\documentclass[prb,aps,twocolumn,groupedaddress,floats,showpacs,final,superscriptaddress]{revtex4-2}
\usepackage[latin3]{inputenc}
\usepackage[makeroom]{cancel}
\usepackage{graphicx}
\usepackage{amsmath}
\usepackage{amsfonts}
\usepackage{amssymb}
\usepackage{chngcntr}
\usepackage{color}
\usepackage{cancel}
\usepackage[left=2cm,right=2cm,top=2cm,bottom=2cm]{geometry}

\usepackage{graphicx}
\usepackage{dcolumn}
\usepackage{bm}
\usepackage{simplewick}
\usepackage{array}
\usepackage{appendix}

\RequirePackage[
   hyperindex,colorlinks,bookmarksnumbered,
   plainpages=true,pdfstartview=FitH]{hyperref}
\hypersetup{linkcolor=blue,urlcolor=blue,citecolor=blue}
\usepackage{hyperref}

\definecolor{purple}{rgb}{0.5,0,0.6}

\usepackage{ulem}

\begin{document}

\title{
Interactions-controlled magnetotransport in two-dimensional massless-massive fermion mixtures}

\author{
Y. Huang}
\affiliation{Department of Materials Science and Engineering, Southern University of Science and Technology, 1088 Xueyuan Blvd, Shenzhen, 518055, China}
\affiliation{Department of Physics, Guangdong Technion--Israel Institute of Technology, 241 Daxue Road, Shantou, Guangdong, China, 515063}

\author{D.~S.~Eliseev}
\affiliation{Novosibirsk State Technical University, Novosibirsk 630073, Russia}

\author{V.~M.~Kovalev}
\affiliation{Novosibirsk State Technical University, Novosibirsk 630073, Russia}

\author{O.~V.~Kibis}
\affiliation{Novosibirsk State Technical University, Novosibirsk 630073, Russia}


\author{Yu. Yu. Illarionov}
\affiliation{Department of Materials Science and Engineering, Southern University of Science and Technology, 1088 Xueyuan Blvd, Shenzhen, 518055, China}

\author{I.~G.~Savenko}
\email[Corresponding author: ]{ivan.g.savenko@gmail.com}
\affiliation{Department of Physics, Guangdong Technion--Israel Institute of Technology, 241 Daxue Road, Shantou, Guangdong, China, 515063}
\affiliation{Technion-Israel Institute of Technology, 32000 Haifa, Israel}


\begin{abstract}
The presence of two types of holes, namely the Dirac holes and the massive holes, in a two-dimensional sample exposed to an external permanent magnetic field leads to the emergence of the temperature and magnetic field-dependent contribution to the resistivity due to their interactions.
Taking a HgTe-based two-dimensional semimetal as a testbed, we develop a theoretical model describing the role of interactions between the degenerate massive and massless Dirac particles for the magnetoconductivity and resistivity in the presence of a classical magnetic field.
If only the Dirac holes are present in the system, the magnetoconductivity acquires a finite interaction-induced contribution, which would vanish for the parabolic spectrum.
It demonstrates $T^4\ln(1/T)$ behavior at low temperatures for short-range interhole interaction potential, and $T^2$-like behavior in the case of long-range interhole interaction potential.
However, the magnetoresistivity and the Hall effect are not affected by the Dirac holes interparticle correlations in the lowest order of interparticle interaction.
In contrast to this, the presence of two types of holes provides a finite contribution to the magnetoconductivity, magnetoresistivity, and the classical Hall effect resistivity.
The temperature behavior of the magnetoconductivity here is $\sim T^2$ in the case of the short-range constant interparticle interaction potential and $T^2\ln(1/T)$ for the bare unscreened Coulomb interaction.
A classically strong magnetic field suppresses the interaction-induced corrections to magnetoresistivity of massless-massive hole gas mixture.
\end{abstract}

\maketitle

\section{Introduction}
In conventional semiconductors with a parabolic spectrum of electrons and holes, particle-particle interaction does not usually impact the electric current density due to the Galilean invariance of the system and the subsequent self-compensation of the Umklapp scattering processes.
Thus, the conductivity is only determined by impurities at low temperatures and phonons at higher temperatures, and the Coulomb interaction between the carriers of charge can be disregarded.

However, recent technological progress allowed for the fabrication of clean, purely two-dimensional (2D) semiconducting materials, in which there can emerge the third regime of operation at low temperatures: the regime, in which the interparticle collisions can become dominating over the impurities and phonon-mediated scattering.
An example is nondegenerate electron gas in monolayer and bilayer graphene.
The electron-hole pairs appear there as a result of thermal excitation~\cite{NamNatPhys2017, doi:10.1126/sciadv.abi8481}, and their collisions give a finite contriburtion to conductivity.
Since the collision rate is linear in $T$, the resistivity is proportional to $T^2$~\cite{DEHAAS1934609, PhysRevLett.129.206802}.
In transition metal dichalcogenides, the intervalley scattering also provides a nonvanishing correction to conductivity due to the interparticle interactions~\cite{PhysRevB.109.245414,
PhysRevB.110.L041301}.

Another example of a system with similar properties is a semimetal such as a HgTe-based quantum well (QW) hosting two types of carriers of
charge (electrons and holes) with different
dispersions, thus allowing for the breaking of the Galiliean
invariance when the density of both the gases of particles is sufficient (degenerate electron and hole gases regime)~\cite{PhysRevB.67.115316, PhysRevB.102.155411, PhysRevB.106.085411}.
As a result, there emerges a friction between electrons and holes, which leads to the finite contribution to resistivity~\cite{OlshanetskyJETP2009}.

In general, HgTe-based materials constitute a platform, which is exceptionally rich since they are characterized by two critical parameters~\cite{doi:10.1126/science.1133734}, which define the on-going physics~\cite{PhysRevLett.95.146802, PhysRevLett.95.226801, PhysRevLett.96.106802}.
The first parameter is fixed at the growing phase: it is the QW width.
It defines the spectrum of the QW particles~\cite{doi:10.1126/science.1133734}.
At certain width (around 6 nm), the band gap can even disappear.
If the QW width is in the vicinity of this critical width, the material turns into a 2D topological insulator~\cite{doi:10.1126/science.1148047, GUSEV2019113701} or a 2D semimetal~\cite{EntinJETP2013}.
Such a regime is manifested by the presence of a single valley representing a Dirac cone~\cite{ButtnerNatPhys2011} with rather unique transport properties, including the magnetotransport in classical magnetic fields~\cite{KvonJETPL2008}, cyclotron resonance~\cite{nano12142492}, weak localization phenomena~\cite{KozlovHETPL2013}, and the quantum Hall regime \cite{PhysRevB.96.045304}.
\begin{figure*}[t!]
\includegraphics[width=1.7\columnwidth]{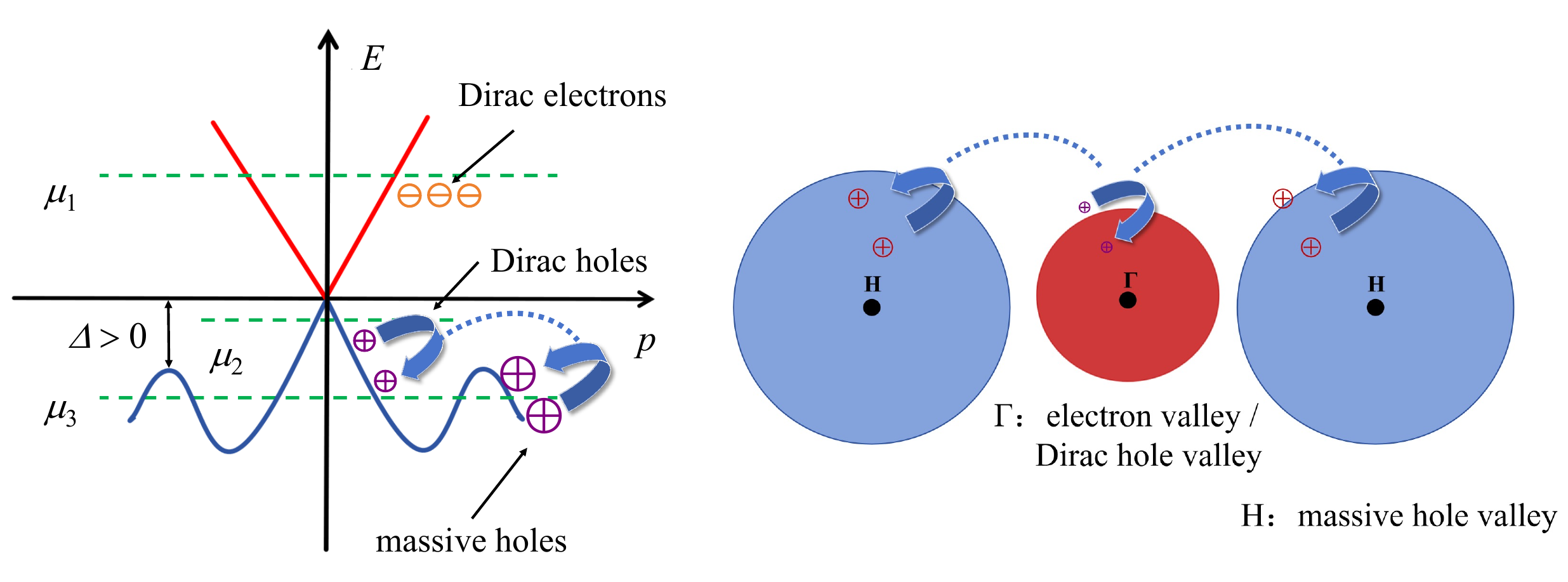}
\caption{Band structure of a 2D gas containing Dirac particles and massive particles.
Red domain of the dispersion corresponds to the electron-like excitations, whereas the blue domain indicates the hole-like excitations of the system.
Green (dashed) horizontal lines indicate three different regimes under study.
The chemical potentials $\mu_1$ and $\mu_2$ correspond to the exciting of only degenerate massless electron or hole Fermi gas, respectively.
The position of $\mu_3$ represents a degenerate massive-massless Fermi gas mixture (given $\mu_3-\Delta\gg T$).
The right-hand side panel shows a schematic of relative positions of two hole and electron valleys: the blue circles correspond to the holes with a quadratic dispersion, and the red one is either for the Dirac electrons or Dirac holes with a linear dispersion.
}
\label{Fig1}
\end{figure*}

The second critical parameter can be in-situ controlled: it is the gate voltage.
Depending on the gate voltage, it is possible to achieve different transport regimes: from a topological insulator to a semimetal.
Moreover, it is possible to observe a switch between different regimes of operation since the gate controls the density of the carriers of charge, and thus, the degeneracy factor.

There also exists another control parameter which not only serves as an alternative tool to change the system behavior, but allows for the study of new effects (with various temperature dependencies): the magnetic field.
The violation of various symmetries in the presence of a magnetic field may result in unusual transport phenomena in 1D nanostructuries caused by the electron-electron~\cite{Kibis_1992} and electron-phonon interaction~\cite{Kibis_2001, Kibis_2002}.
An in-plane magnetic field in a quasi-2D structure may result in the asymmetry of electron-phonon interaction under the action of surface acoustic wave or uniform heating~\cite{Kibis_1999}.
Experimental manifestation of specific transport phenomena occurring under the violation of the time-reversal symmetry was reported in the hybrid semiconductor-ferromagnetic structures~\cite{Lawton-2002}.

Furthermore, particle-particle collisions in systems with parabolic dispersion of carriers does not directly affect the transport properties due to the presence of the Galilean invariance. Nevertheless, the latter symmetry may be violated in quasi-2D structures with several subbands, thus, resulting in a sufficient influence of interparticle scattering from different subbands on the total electric current~\cite{PhysRevB.59.2376, RefHoleHole}.

Recently, it has been experimentally shown that HgTe-based QWs demonstrate a $T^2$-scaled conductivity at low temperatures~\cite{PhysRevLett.134.196303}.
This temperature behavior was associated with the interaction-induced correction to the Drude conductivity of degenerate Dirac electrons with a linear dispersion even being single-band.
The theory of electron-electron interaction-induced correction to conductivity supports these experimental findings~\cite{PhysRevLett.134.196303, lnn9-blkj}.

Current paper aims at an extension of this theory considering an exposed to magnetic field HgTe-based QW hosting the carriers of charge possessing the energy dispersion depicted in Fig.~\ref{Fig1}.
It consists of Dirac massless electrons and holes sector accompanied with the hole valleys with a massive dispersion (in HgTe structures, the massive hole mass is $m\approx0.15m_e$ with $m_e$ the free electron mass).
Such a spectrum is typical for HgTe QW structures with two kinds of holes coexisting at certain QW width near 6~nm.
Then, in addition to the Dirac holes with the spectrum centered at the $\Gamma$-point of the Brillouin zone, there also exist massive holes in the neighboring valleys.
Below we analyze the effect of interparticle interaction to the magentoconductivity, magnetoresistivity, and the Hall effect in such structures.
We demonstrate that the interparticle interaction between massless electrons (or massless holes) produce the temperature and magnetic field corrections to the magnetoconductivity but no corrections to the magnetoresistivity.
In contrast, the nonzero corrections to magnetoresistivity is produced by the interaction between massless and massive holes.
Thus, we explore the interaction-induced magnetoresistivity, where the interaction is due to the mutual presence of the particles with nonrelativistic and quasirelativistic spectra.


\section{Magnetoconductivity of interacting hole gas}
At low temperatures, when the carriers of charge are degenerate, the system represents a degenerate massless electron Fermi gas in the n-doped regime, whereas in the p-doped regime, when the Fermi level $\mu>\Delta$ (in the hole representation), the system constitutes a mixture of degenerate massive and massless hole Fermi gases (Fig.~\ref{Fig1}).
If $\Delta>\mu>0$, the system is a Fermi gas of only massless  Dirac holes.
We will study the interaction-mediated corrections to magnetotransport properties of the system in different regimes.

We assume that the residual conductivity at small (zero) temperature is determined by the particle-impurity scattering.
In contrast, the low-temperature corrections to the system conductivity mainly stem from particle-particle scattering processes.
The term ``low temperatures'' in this sense means that the electron-electron (and hole-hole) scattering times are much larger than the particle-impurity ones, and the particle-phonon scattering processes are neglected.
Indeed, at low enough temperatures and in a degenerate-gas, the particle-particle scattering time usually behaves as $1/\tau_{pp}\sim T^2/\mu$.
Hence, the condition $\tau_{pp}\gg\tau$ (where $\tau$ is the corresponding particle momentum relaxation time on impurities) allows us to describe particle-particle scattering perturbatively withing the Boltzmann transport equation approach~\cite{RefPal}, employing the method of successive approximations using the smallness of the particle-particle collision integral.
An exact inequality in the case of a Dirac dispersions reads $\hbar/\tau_{pp}=(e^2/\epsilon\hbar v)^2T^2/\mu\ll\hbar/\tau$.
This gives an estimation for the upper limit of temperature around 90~K for $\tau=100$~fs.
However, this is an overestimation.
In experiments, the realistic range of temperatures, where interparticle interaction play important role, is 30--50~K.

At the same time, we assume a classical magnetic field with arbitrary magnitude, thus, the parameter $\omega_c\tau$ (with $\omega_c$ the cyclotron frequency) can acquire arbitrary values for both massless and massive holes.

In the following two subsections, we consider the magnetotransport in the p-doped regime for Fermi gas of only Dirac holes (which is analogous to the Fermi gas of  Dirac electrons) and a massless-massive holes mixture.


\subsection{Magnetoconductivity of the gas of interacting massless hole particles}
The general Boltzmann equation describing the scattering of massless holes on impurities and other holes reads as (here and below we use $c=\hbar=1$ units)~\cite{PhysRevB.109.245414}
\begin{eqnarray}\label{nregime1}
e({\bf E}+[{\bf v}_{\bf p}\times{\bf B}])
\cdot\nabla_{\bf p}f_{\bf p}+\frac{f_{\bf p}-n_{\bf p}}{\tau}=Q\{f_{\bf p}\},
\end{eqnarray}
where $e>0$ is a hole charge,  $\mathbf{E}$ and $\mathbf{B}$ are the electric and magnetic fields, $\mathbf{v}_\mathbf{p}=v\mathbf{p}/p=v(\cos\phi_\mathbf{p},\sin\phi_\mathbf{p})$ is a velocity of massless holes with a dispersion $\varepsilon_{\bf p}=vp$, $f_\mathbf{p}$ and $n_\mathbf{p}$ are their non-equilibrium and equilibrium distribution functions, respectively; $\tau$ is a corresponding impurity-scattering time that we assume to be independent on the hole energy, and $Q\{f_{\bf p}\}$ is the Coulomb hole-hole collision integral.
Furthermore, the non-equilibrium distribution function can be written as a sum of zero-order $\delta f_{\bf p}$ (determined only by scattering on impurities) and first-order $\delta f^{C}_{\bf p}$ contributions with respect to h-h collision integral: $f_{\bf p}-n_{\bf p}=\delta f_{\bf p}+\delta f^{C}_{\bf p}$.
These functions are the solutions of the following linear-in-${\bf E}$ equations written in the polar coordinate system in the $\bf p$-space:
\begin{gather}\label{nregime2}
\left(
\frac{\partial}{\partial\phi_{\bf p}}
-\frac{1}{\omega_c\tau}
\right)
\delta f_{\bf p}=
\frac{e}{\omega_c}({\bf E}\cdot{\bf v}_{\bf p})n_{\bf p}',\\\nonumber
\left(
\frac{\partial}{\partial\phi_{\bf p}}
-\frac{1}{\omega_c\tau}
\right)
\delta f^C_{\bf p}=
-\frac{Q\{\delta f_{\bf p}\}}{\omega_c},
\end{gather}
where $\omega_c=evB/p$ is the cyclotron frequency of Dirac holes, and prime means the derivative with respect to hole energy $\varepsilon_{\bf p}$.
The general solutions of these equations read as
\begin{gather}\label{nregime3}
\delta f_{\bf p}=
-e\tau n_{\bf p}'
\int\limits_0^\infty
\frac{d\xi}{\omega_c\tau}
e^{-\xi/\omega_c\tau}({\bf E}\cdot{\bf v}_{\bf p})_{\phi_{\bf p}\rightarrow\phi_{\bf p}+\xi},\\\nonumber
\delta f^C_{\bf p}=
\tau\int\limits_0^\infty
\frac{d\xi}{\omega_c\tau}
e^{-\xi/\omega_c\tau}Q\{\delta f_{\bf p}\}_{\phi_{\bf p}\rightarrow\phi_{\bf p}+\xi}.
\end{gather}
The integration in the first expression in Eq.\eqref{nregime3} yields
\begin{gather}\label{nregime4}
\delta f_{\bf p}=
-e\tau E_xn_{\bf p}'\frac{v_x({\bf p})-\omega_c\tau v_y({\bf p})}{1+\omega_c^2\tau^2}\equiv\chi_{\bf p}n_{\bf p}'.
\end{gather}
Next, we can substitute this expression in the second expression in Eq.~\eqref{nregime3}, providing the interaction-induced correction to the hole distribution function.
The interaction-induced correction to current density can be found using the second expression in Eq.~\eqref{nregime3} as
\begin{gather}\label{nregime5}
{\bf j}=e\sum_{{\bf p}}{\bf v}({\bf p})\delta f^C_{{\bf p}},
\end{gather}
\textcolor{black}{where we imply the summation over the spins and valleys.}
To find an explicit expression for the current density, we should specify the hole-hole collision integral, $Q\{\delta f_{\bf p}\}$.
Initially, it has a standard nonlinear form.
In the framework of our assumptions, it can be linearized with respect to the electric field or the hole distribution function $\delta f_{\bf p}$, which is linear in ${\bf E}$.
We find:
\begin{gather}\nonumber
Q\{\delta f_{\bf p}\}=-2\pi\sum_{\mathbf{k}',\mathbf{p}',\mathbf{q}}
|U_{\mathbf{p}'-\mathbf{p}}|^2(\chi_\mathbf{p}-\chi_{\mathbf{p}'}+\chi_\mathbf{k}-\chi_{\mathbf{k}'})
\\
\label{nregime6}
\times(n_\mathbf{p}-n_{\mathbf{p}'})
(n_\mathbf{k}-n_{\mathbf{k}'})
\delta_{\mathbf{k}',\mathbf{k}+\mathbf{q}}
\delta_{\mathbf{p}',\mathbf{p}-\mathbf{q}}\\
\nonumber
\times\int\limits_{-\infty}^{\infty}\frac{d\omega}{4T\sinh^2(\omega/2T)}
\delta(\varepsilon_{{\bf k}'}-\varepsilon_{\bf k}-\omega)
\delta(\varepsilon_{{\bf p}'}-\varepsilon_{\bf p}+\omega).
\end{gather}
Here, ${\bf q}$ and $\omega$ are the momentum and energy transferred between holes in the collision event.
For strongly degenerate holes (the case we consider), $\omega$ is much smaller than the Fermi energy, and the expansions $n_{\bf p}-n_{{\bf p}'}\approx \omega n_{\bf p}'$, $n_{\bf k}-n_{{\bf k}'}\approx -\omega n_{\bf p}'$ are substantiated.
Taking into account these relations and integrating over ${\bf p}'$ and ${\bf k}'$ in Eq.~\eqref{nregime6}, gives the current density correction
\begin{gather}\label{nregime7}
{\bf j}=-2\pi(e\tau)^2E_x\sum_{{\bf p},{\bf k},{\bf q}}|U_{\bf q}|^2
\int\limits_0^\infty
\frac{d\xi}{\omega_c\tau}
e^{-\xi/\omega_c\tau}
n_{\bf p}'n_{\bf k}'
\\\nonumber
\times\Biggl[{\bf v}({\bf p})\frac{\Delta v_{x}({\bf p},{\bf k},{\bf q})-\omega_c\tau\Delta v_{y}({\bf p},{\bf k},{\bf q})}{1+\omega_c^2\tau^2}
\int\limits_{-\infty}^{\infty}\frac{\omega^2d\omega}{4T\sinh^2(\frac{\omega}{2T})}
\\\nonumber
\times\delta(\varepsilon_{{\bf k}+{\bf q}}-\varepsilon_{\bf k}-\omega)
\delta(\varepsilon_{{\bf p}-{\bf q}}-\varepsilon_{\bf p}+\omega)
\Biggr]_{\phi_{\bf p}\rightarrow\phi_{\bf p}+\xi},
\end{gather}
where $\Delta v_{x(y)}=v_{x(y)}({\bf p})-v_{x(y)}({\bf p}-{\bf q})+v_{x(y)}({\bf k})-v_{x(y)}({\bf k}+{\bf q})$.

The current from Eq.~\eqref{nregime7} can be found analytically.
However, the procedure is somewhat cumbersome, thus we present it  in the Supplemental Material~\cite{[{See Supplemental Material at [URL], which gives the details of the analysis of all the relevant Feynman diagrams}]SMBG}).
After calculations, the interaction-induced correction to magnetoconductivity tensor reads ($\delta\sigma_{xy}=-\delta\sigma_{yx}, \delta\sigma_{xx}=\delta\sigma_{yy}$):
\begin{gather}\label{nregime8}
\left(\begin{matrix}
\delta\sigma_{xx} \\
\delta\sigma_{xy}
\end{matrix}\right)=
\textcolor{black}{(g_s^Dg_v^D)^2}
\left(\begin{matrix}
1-\omega_c^2\tau^2 \\
2\omega_c\tau
\end{matrix}\right)
\frac{\sigma_0^D(T)}{(1+\omega_c^2\tau^2)^2},~~~\textrm{where}\\
\nonumber
\sigma_0^D(T)=\left(\frac{e\tau}{2\pi v}\right)^2\int\limits_{-\infty}^\infty
\frac{-\omega^4 d\omega}{(4\pi T)\sinh^2(\frac{\omega}{2T})}
\int\limits_{|\omega|/v}^\infty \frac{qdq}{2\pi\varepsilon_q^2}
|U_{\bf q}|^2.
\end{gather}
Here, $g_s^D=2$ and $g_v^D=1$ are spin and valley degeneracy factors for the Dirac holes, $\varepsilon_q=vq$, and $\sigma_0^D(T)$ is the interaction-induced correction to conductivity of Dirac holes at zero magnetic field.
Its temperature dependence is determined by the Fourier component of the interhole interaction potential $U_q$.

In HgTe-based systems, an external gate usually controls the density of charge carriers. This density directly defines the screening effects. Thus, sometimes the short-range, and sometimes the actual Coulomb potential suits better for the analysis of the experimental data.
Thus, we present here both the models and
consider two fundamentally important limiting cases.
The first one is the short-range interparticle interaction potential, $U_{\bf q}=U_0$=const, and the second is the long-range interaction (the bare Coulomb potential), $U_{\bf q}=2\pi e^2/\epsilon q$.
In the former case, the upper limit in the integral should be taken as $2 p_F=2\mu/v$. Thus, a general expression Eq.~\eqref{nregime8} yields (restoring $\hbar$):

\begin{figure}[t!]
\includegraphics[width=1.03\columnwidth]{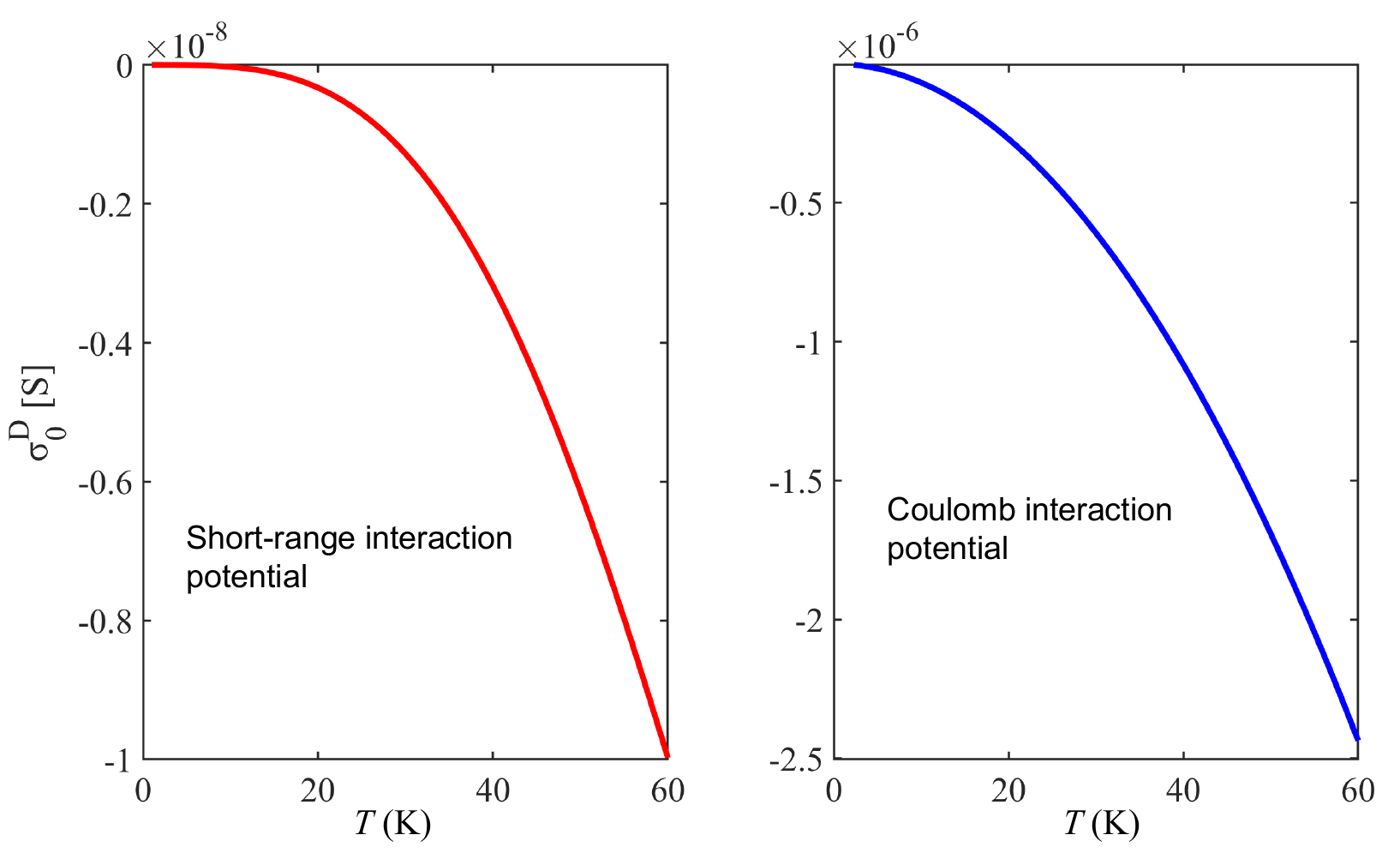}
\caption{Temperature dependence of the correction to the massless Dirac hole conductivity.
Left panel: short-range interaction potential between holes, $\sigma_0^D$ exhibits a $T^4 \ln(\mu/T)$ behavior; right panel: long-range Coulomb potential, $\sigma_0^D\sim T^2$.
The following parameters for a typical HgTe structures were used: Fermi velocity $v = 7 \times 10^5$ m/s, effective mass $m = 0.15m_e$ (with $m_e$ the free electron mass), dielectric constant $\epsilon = 10$, momentum relaxation time $\tau = 100$~fs, and chemical potential $\mu = 10$~meV.}
\label{Fig2}
\end{figure}

\begin{gather}\label{nregime9.1}
\sigma_0^D(T)=-\frac{e^2}{15\hbar}\left(\frac{T\tau}{\hbar}\right)^2\left(\frac{U_0T}{\hbar^2v^2}\right)^2\ln{(\mu/T)}.
\end{gather}
The latter case (long-range interaction) gives
\begin{gather}\label{nregime9.2}
\sigma_0^D(T)=-\frac{e^2}{6\hbar}\left(\frac{T\tau}{\hbar}\right)^2\left(\frac{e^2}{\epsilon\hbar v}\right)^2.
\end{gather}
%
%
%
%
%
Figure~\ref{Fig2} shows the dependencies of~\eqref{nregime9.1} and~\eqref{nregime9.2} on temperature, thus we can compare $\sigma^D_0(T)$ for both short-range and long-range interaction models.
Evidently, the latter produces a stronger interaction correction to conductivity at low temperatures.
The short-range potential in Eq.~\eqref{nregime9.1} was estimated as the overscreened Coulomb potential,  $U_0=2\pi e^2/\epsilon q_s$, where $q_s$ is a screening wave vector for degenerate Dirac particles.
Other relevant parameters used in the calculations are mentioned in the figure caption.

It should be noted that expressions~\eqref{nregime9.1} and~\eqref{nregime9.2} were derived for the case of Fermi gas of only Dirac holes.
A similar expression obviously holds for the degenerate Dirac Fermi gas of electrons, corresponding to the position of the chemical potential $\mu=\mu_1$ in Fig.~\ref{Fig1}.

\begin{figure}[t!]
\includegraphics[width=1.03\columnwidth]{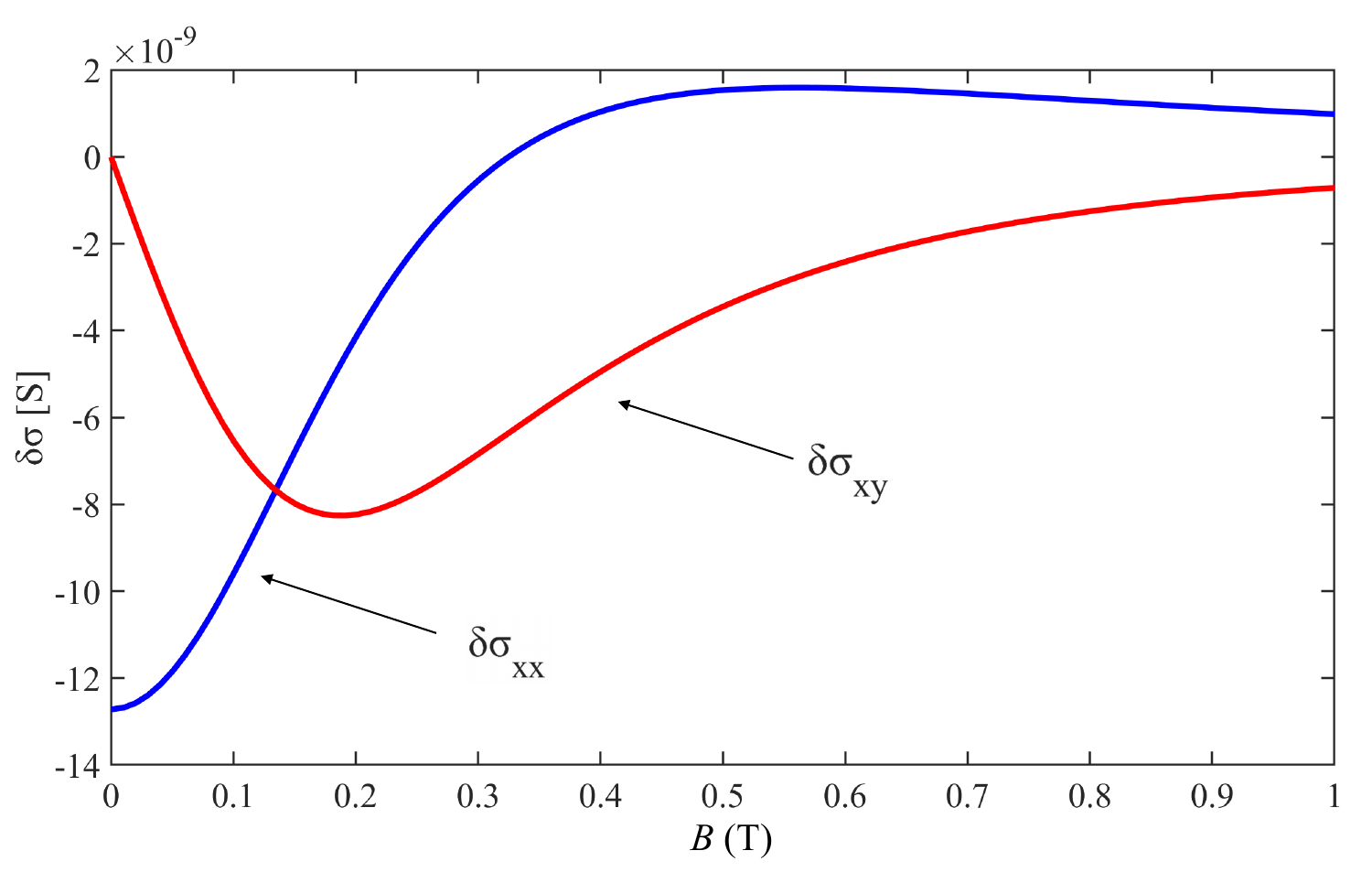}
\caption{The massless Dirac hole longitudinal magnetoconductivity \(\delta\sigma_{xx}\) and transverse magnetoconductivity \(\delta\sigma_{xy}\) as functions of the magnetic field for the case of short-range interaction potential.
The following parameters were used in the calculations: \(T=40\) K, Dirac hole density \(N_d = 1.9 \times 10^{14}\) cm\(^{-2}\), massive hole density \(N_h = 3.2 \times 10^{15}\) cm\(^{-2}\).
Other parameters are the same as in Fig.~\ref{Fig2}}
\label{Fig3}
\end{figure}

Figure~\ref{Fig3}  shows the dependencies of longitudinal and transverse (Hall) magnetoconductivities on magnetic field.
They are provided by Eq.~\eqref{nregime8} with $\sigma^D_0(T)$ taken for a short-range interaction potential (Eq.~\eqref{nregime9.1}).
The interaction-induced corrections behave non-monotonously.
Moreover, if the transverse conductivity $\delta\sigma_{xy}$ is always negative, the longitudinal one, $\delta\sigma_{xx}$, changes its sign, which is reflected in the factor $1-\omega^2_c\tau^2$ in the formula.
This factor is a universal feature, and it is well known in the systems with parabolic dispersions (see, e.g.,~\cite{PhysRevB.25.2196}).
In the regime of classically large magnetic fields, $\omega_c\tau\gg1$, both corrections vanish.

\subsection{Magnetoconductivity of the interacting massless-massive hole gas mixture}
Next, let us consider an interacting mixture of massive and massless holes (that corresponds to $\mu=\mu_3$ in Fig.~\eqref{Fig1}).
Three types of inter-particle interactions are present here: the scattering in (i) massive and (ii) massless hole subsystems individually and (iii) between holes from different subsystems.
The interaction between massive holes has no effect on the conductivity (as we discussed in the introduction).
The interaction between massless holes was scrutinized in the previous subsection.
Hence, let us address the interaction between different types of holes.

This correction, in turn, splits into two contributions: massless holes scatter off the massive holes and vice versa.
We will disregard the possible processes of conversion of one type of holes to another because it would be accompanied by a large momentum transfer (between the massive and massless hole valleys).
Obviously, such processes are suppressed.
We will also disregard the interband scattering processes, in which the holes can be converted into massless electrons, thus, assuming $|\mu_3|\gg T$.

Given the restrictions discussed above, the density of each component of the hole mixture is conserved, and we can write two separate Boltzmann equations, describing the dynamics of each component exposed to an external static electric field ${\bf E}=(E_x,0)$ independently.
Their coupling is due to the Coulomb scattering integral:
\begin{gather}\nonumber
e[\mathbf{E}+\mathbf{v}_\mathbf{p}\times\mathbf{B}]\cdot\nabla_\mathbf{p}f_\mathbf{p}=-\frac{f_\mathbf{p}-n_\mathbf{p}}{\tau_{p}}+Q_\mathbf{p}\left\{f_\mathbf{p},f_\mathbf{k}
\right\},\\
e[\mathbf{E}+\mathbf{v}_\mathbf{k}\times\mathbf{B}]\cdot\nabla_\mathbf{k}f_\mathbf{k}=-\frac{f_\mathbf{k}-n_\mathbf{k}}{\tau_{k}}+Q_\mathbf{k}\left\{f_\mathbf{p},f_\mathbf{k}
\right\},
\end{gather}
where $\tau_p$ and $\tau_k$ are the impurity scattering times (we used $\tau$ instead of $\tau_p$ in the previous subsection), and $Q_\mathbf{p}$ and $Q_\mathbf{k}$ are the Coulomb interaction-induced scattering integrals.
Here and in what follows, the index `p' and momentum $p$ are related to massless Dirac holes with energy $\varepsilon_p=vp$, whereas the index `k' and momentum $k$ are associated with the massive holes characterized by the parabolic dispersion: $\varepsilon_k=k^2/2m+\Delta$; all the energies are counted from the Dirac point.
The cyclotron frequencies are $\omega_p=eBv^2/\mu$ for Dirac holes and $\omega_k=eB/m$ for massive holes.

The procedure of finding the Boltzmann equations' solutions is identical to the case of Fermi gas of only Dirac holes considered in the previous subsection (see the details in the Supplemental Material~\cite{[{See Supplemental Material at [URL], which gives the details of the derivations}]SMBG}).
After some calculus, we come up with the analytical but cumbersome expressions for current densities (interaction-induced corrections), see Appendix A, formulas~\eqref{EqMain14}-\eqref{EqMain17}.

Further analytical analysis of these formulas is only possible in particular limiting cases.
First, let us take the model of momentum-independent hole-hole interaction potential: $U_{\bf q}=U_0$.
This model is applicable either in the case of a formally arbitrary contact interaction or in the case of a strongly screened interaction system with $U_0\approx2\pi e^2/\epsilon q_s$, where $q_s$ is the screening wavevector. The main contribution to the screening effect comes from the massive holes due to large value of their effective mass and, in turn, the  density of states.
Thus, we take the screening wave vector in the form $q_s=2/a_B$, where $a_B=\hbar^2\epsilon/e^2m$ is the Bohr radius of massive holes.
An estimation gives $q_s\approx 5.7\cdot 10^6$~cm$^{-1}$ for $\epsilon=10$ and $m=0.15 m_0$.
\begin{figure}[t!]
\centering
\includegraphics[width=1.04\columnwidth]{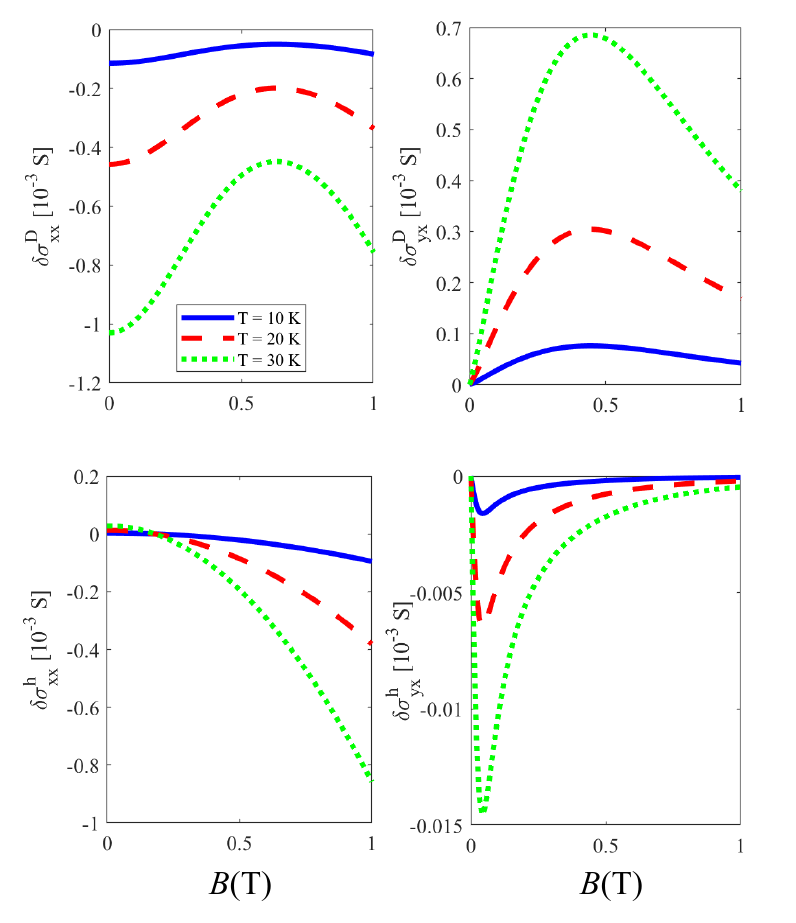}
\caption{The magnetoelectric conductivity correction for massless holes $\delta\sigma_{xx,yx}^{D}$ due to scattering massless holes off massive ones (top panels) and massive holes $\delta\sigma_{xx,yx}^{h}$ due to the scattering massive holes off massless ones (bottom panels) as functions of the magnetic field at different temperatures: 10~K (blue), 20~K (red), and 30~K (green).
Here, $\beta=\tau_p m v/(\tau_k p_0)$ characterizes the ratio of the relaxation times and velocities for the holes, and $\tilde g=g_{s}^{D} g_{v}^{D} g_{s}^{h} g_{v}^{h}=8$, with $g_{s}^{h}=2$ and $g_{v}^{h}=2$.
Parameters: the Dirac hole density and the massive hole density are the same as in Fig.~\ref{Fig3}; Dirac hole mobility $\mu_d=24$~m$^2$/V$\cdot$s; massive hole mobility $\mu_h=1.3$~m$^2$/V$\cdot$s;  the short-range (contact) interaction potential $U_0 = 2\pi e^2/(\epsilon q_s)$ with the screening wave-vector $q_s$. Other parameters are the same as in previous plots.}
\label{Fig4}
\end{figure}
\begin{figure}[t!]
\centering
\includegraphics[width=1.03\columnwidth]{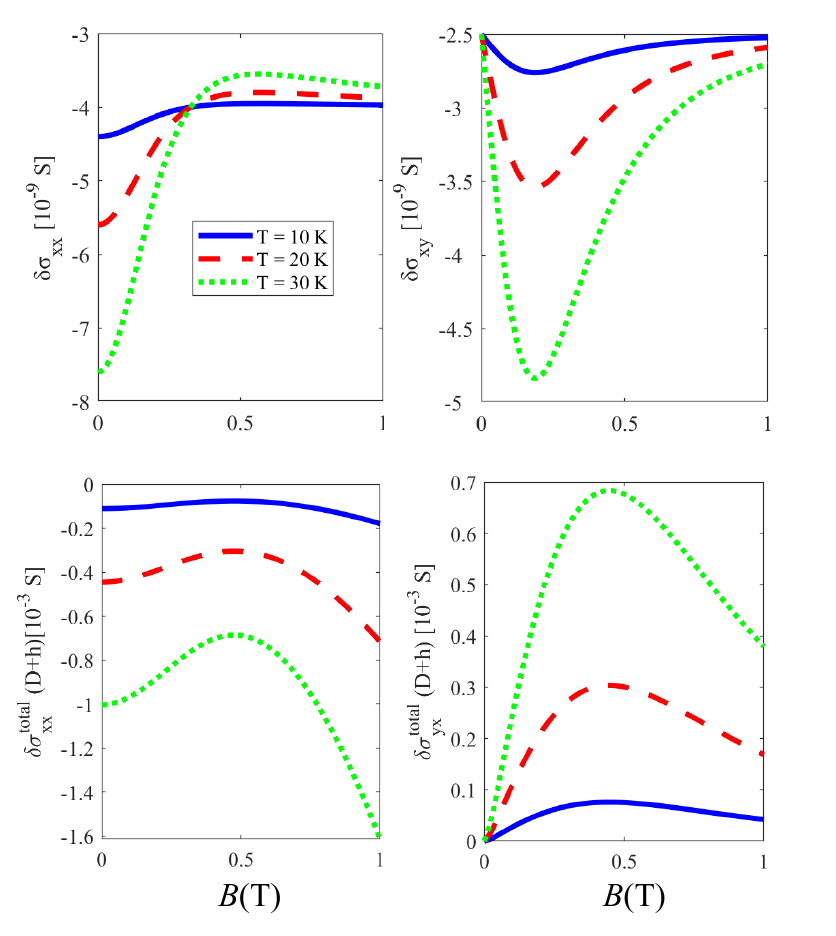}
\caption{Comparison of interaction-induced  corrections to magnetoconductivity for the (i) pure Dirac hole gas (upper panels, Eq.~\eqref{nregime8}), and (ii) massive-massless hole gas mixture (lower panels, sum of Eqs.~\eqref{masslessholes} and Eqs.~\eqref{massiveholes}) for the temperatures 10~K (blue), 20~K (red), and 30~K (green) calculated for short-range interhole interacting potential.}
\label{Fig5}
\end{figure}

Reasonably assuming $\omega,q^2/2m,\varepsilon\ll\mu$ under the square roots in Eqs.~\eqref{EqMain14}-\eqref{EqMain17}, we find that the lower limit in $q-$integrals is $q=|\omega|/v$, and the upper limit is $\textrm{min}[2p_0, 2k_0v]$, where $p_0^2=4\pi\hbar^2N_d/\textcolor{black}{g^D_{s}g^D_{v}}$, $4\pi\hbar^2 N_h/\textcolor{black}{g^h_{s}g^h_{v}}=k_0^2$ with $N_{d(h)}$ being the Dirac and massive hole densities.
Taking $N_d<N_h$, we find the expression for the Dirac holes' contribution:
\begin{gather}\label{masslessholes}
\delta\sigma^D_{xx}=\sigma_{0}(T)\textcolor{black}{\tilde g}
\left(\beta\eta_{pp}-\eta_{kp}\right),
\\\nonumber
\delta\sigma^D_{yx}
=\sigma_{0}(T)\textcolor{black}{\tilde g}
\left(\beta\zeta_{pp}-\zeta_{kp}\right),
\end{gather}
where $\beta=\tau_pmv/\tau_kp_0$.
The massive holes' contribution reads:
\begin{gather}\label{massiveholes}
\delta\sigma^h_{xx}
=\sigma_{0}(T)\textcolor{black}{\tilde g}
\left(\beta^{-1}\eta_{kk}-\eta_{kp}\right),
\\\nonumber
\delta\sigma^h_{yx}
=\sigma_{0}(T)\textcolor{black}{\tilde g}
\left(\beta^{-1}\zeta_{kk}-\zeta_{kp}\right),
\end{gather}
where the factors $\eta_{kp}$ and $\zeta_{kp}$ contain the dependencies on the external magnetic field:
\begin{eqnarray}
\eta_{kp}=\frac{\tau_k\omega_k\tau_p\omega_p-1}{(1+\tau_k^2\omega_k^2)(1+\tau_p^2\omega_p^2)},
\\
\zeta_{kp}=\frac{\tau_k\omega_k+\tau_p\omega_p}{(1+\tau_k^2\omega_k^2)(1+\tau_p^2\omega_p^2)},
\end{eqnarray}
and
\begin{gather}\label{temperatureHHcond}
\sigma_{0}(T)=\frac{e^2}{3\hbar}
\left(\frac{\tau_p\tau_k T^2}{\hbar^2}\right)
\left(\frac{m U_0}{2\pi\hbar^2}\right)^2
\left(\frac{p_0^2}{k_0mv}\right)
\end{gather}
is the interaction-mediated conductivity in the absence of an external magnetic field.
It contains all the temperature behavior of all the conductivities~\eqref{masslessholes} and~\eqref{massiveholes}.
%
%
\begin{center}
\begin{table*}
\begin{tabular}{ |p{3cm}||p{5cm}|p{6cm}|  }
 \hline
 \multicolumn{3}{|c|}{Different interaction cases} \\
 \hline
 Type of interaction & Contact interaction, $U_{\bf q}=U_0$       & Coulomb interaction, $U_{\bf q}=2\pi e^2/\epsilon q$\\
 \hline
 massless-massless   & \,\,\,\,$\sim T^4\ln(\mu/T)$     & \,\,\,\,$\sim T^2$\\
 massive-massless    &  \,\,\,\,$\sim T^2$            & \,\,\,\,$\sim T^2\ln(2\mu/T)$\\
 \hline
 \end{tabular}
 \caption{Temperature dependencies of interaction-induced conductivity corrections at $B=0$ due to massless hole scattering, $\sigma_0^D(T)$, and massless-massive hole scattering, $\sigma_0(T)$, processes for long-range and short-range interaction potentials.}
 \end{table*}
\end{center}


So far, we have considered constant Coulomb interaction potential.
In the case of the bare Coulomb potential $U_{\bf q}=2\pi e^2/\epsilon q$ (which often takes place in different systems), the dependence on $T$ in Eq.~\eqref{temperatureHHcond} is different: instead of $\sigma_0(T)\sim T^2$, we find $\sigma_0(T)\sim T^2\ln(2\mu/T)$. All the temperature dependencies of the interaction-induced corrections to the conductivity are summarized in Tab.1 for the comparison.

Figure \ref{Fig4} shows the comparisong of the contributions given by Eq.~\eqref{masslessholes} and Eq.~\eqref{massiveholes} at different temperatures (for short-range interaction, Eq.~\eqref{temperatureHHcond}).
Evidently, the contribution of massive holes (scattered off the Dirac ones) is weaker in comparison with the contribution of Dirac holes scattered off the massive ones.
This is due to the large factor $\beta\gg1$ for reasonable parameters of a HgTe structures determined by the large value of the mobility of the Dirac holes in real HgTe samples.
Thus, the total conductivity correction, which is a sum of Eq.~\eqref{masslessholes} and Eq.~\eqref{massiveholes}, is determined mainly by the scattering of Dirac holes on the massive holes.

Furthermore, let us compare the interaction-induced conductivity of the Fermi gas of Dirac holes only and the conductivity of the mixture (Fig.~\ref{Fig5}).
The two upper panels are the longitudinal and transverse corrections to conductivity of the Dirac hole gas (Eqs.~\eqref{nregime8} and~\eqref{nregime9.1}), whereas the lower panels depict the conductivity of the massless-massive mixture (the sum of Eqs.~\eqref{masslessholes} and~\eqref{massiveholes}).
For comparison, we used the short-range hole-hole interaction potential in both the cases and the same set of temperatures: $T=10$~K, $T=20$~K, and $T=30$~K.
Evidetly, the mixture scattering channel is more effective: it produces several orders of magnitude larger contribution to the magnetoconductivity.
Thus, for the parameters we used, the dominating contribution to the magnetoconductivity stems from the scattering of massless holes off massive ones.

\section{magnetoresistivity of interacting hole gas}

In the previous section, we focused on magnetoconductivity.
The magnetoresistivity with account of interaction-induced corrections can be found as $\hat\rho=\hat\sigma^{-1}$, yielding:
\begin{gather}
\label{MRs}
\rho_{xx}(B)=\frac{\sigma^{0}_{xx}(B)+\delta\sigma_{xx}(B)}
{[\sigma^{0}_{xx}(B)+\delta\sigma_{xx}(B)]^2+[\sigma^{0}_{yx}(B)+\delta\sigma_{yx}(B)]^2},
\end{gather}
where $\sigma^{0}_{\alpha\beta}(B)$ is a bare (interaction-independent) Drude magnetoconductivity tensor.
To find the analytical expression for the magnetoresistivity, we expand Eq.~\eqref{MRs} up to the linear order with respect to the interaction-induced magnetoconductivity terms $\delta\sigma_{\alpha\beta}(B)=\delta\sigma^D_{\alpha\beta}(B)+\delta\sigma^h_{\alpha\beta}(B)$: $\rho_{xx}(B)=\rho^0_{xx}(B)+\delta\rho_{xx}(B)$, where
\begin{eqnarray}\label{MRs1}
&&\rho^0_{xx}(B)=\frac{\sigma^0_{xx}(B)}
{[\sigma^0_{xx}(B)]^2+[\sigma^0_{yx}(B)]^2},
\\\nonumber
&&\delta\rho_{xx}(B)=
-[\rho^0_{xx}(B)]^2\\
\nonumber
&&~~~\times
\left\{\left[1-\left(\frac{\sigma^0_{yx}(B)}{\sigma_{xx}^0(B)}\right)^2\right]
\delta\sigma_{xx}(B)
+2\frac{\sigma^0_{yx}(B)}{\sigma_{xx}^0(B)}\delta\sigma_{yx}(B)\right\}.
\end{eqnarray}
Now, we should consider the cases of massless holes gas and the mixture separately.


\subsection{Magnetoresistivity of the gas of interacting massless hole particles}

Let us analyze the magnetoresistance of interacting massless hole gas.
As we know from the analysis described above, the interaction-induced corrections to the magnetoconductivity strongly depend on the magnetic field.
Despite this, magnetoresistance vanished. Indeed, substituting bare Drude magnetoconductivity of massless holes, $\sigma^0_{xx}(B)=\sigma^0(1+\omega_c^2\tau_p^2)^{-1}$ and $\sigma^0_{xy}(B)=\sigma^0\omega_c\tau(1+\omega_c^2\tau_p^2)^{-1}$ with $\sigma^0$ the Drude conductivity, and the interaction-induced correction~\eqref{nregime8} in Eq.~\eqref{MRs} yields a vanishing contribution to magentoresistivity.
This conclusion holds at least in the first order with respect to the interparticle interaction probability, provided $\sigma_0(T)\sim|U_{\bf q}|^2$.
Thus, the first-order interaction-induced correction to magnetoresistivity and the Hall coefficient are zero if we do not consider the mixture of massive and massless gases.

It should be noted here, that the next-order corrections $\sim\sigma^2_0(T)\sim|U_{\bf q}|^4$ are finite, but they are expectedly smaller (and also, keeping them would be an excess of precision in our case since the collision integral~\eqref{nregime6} only accounts for the particle-particle collisions in the first order concerning the Coulomb interaction potential $\sim|U_{\bf q}|^2$).


\subsection{Magnetoresistivity of interacting massless-massive hole gas mixture}

In the presence of two types of holes, instead of $\sigma^0_{xx(yx)}(B)$ described in Eqs.~\eqref{MRs} and~\eqref{MRs1}), we should consider $\tilde\sigma^0_{xx(yx)}=\sigma^D_{xx(yx)}(B)+\sigma^h_{xx(yx)}(B)$, where
\begin{gather}\label{MRs2}
\hat{\sigma}^{D(h)}(B)=\frac{\sigma^{D(h)}}{1+\omega_{p(k)}^2\tau_{p(k)}^2}
\left(
\begin{matrix}
1 & \omega_{p(k)}\tau_{p(k)} \\
-\omega_{p(k)}\tau_{p(k)} & 1
\end{matrix}
\right),\\\nonumber
\textrm{with}~~~\sigma^D=\frac{e^2N_d\tau_pv}{p_0}~\textrm{and}~\sigma^h=\frac{e^2N_h\tau_k}{m}.
\end{gather}
The general analysis of Eq.~\eqref{MRs2} for the two-component system provides cumbersome analytical solutions.
Thus, for clarity, we will consider the limit of small magnetic fields: $\omega_{p(k)}\tau_{p(k)}\ll1$, which after some algebra yields
\begin{eqnarray}\label{MRs3}
&&\delta\rho_{xx}(B)\approx
\sigma_0(T)\frac{(\beta-1)^2}{\beta(\sigma^D+\sigma^h)^2}
\\\nonumber
&&~~~~~~~~~~~~~~~~~
-\sigma_0(T)\frac{\sigma^D+\beta\sigma^h}{\beta(\sigma^D+\sigma^h)^4}
(\omega_p\tau_p-\omega_k\tau_k)
\\
\nonumber
&&~~~~~~~~~~~~~~~~~~~~~~~~~~~\times
\left\{[\sigma^D+(3\beta-2)\sigma^h]\omega_p\tau_p\right.\\
\nonumber
&&\left.~~~~~~~~~~~~~~~~~~~~~~~~~~~~~~~+[(2\beta-3)\sigma^D-\beta\sigma^h]\omega_k\tau_k\right\}.
\end{eqnarray}
Here, the first line is the interaction-induced correction to the resistivity at zero magnetic field, and the other terms (lines 2-4) reflect the magnetic field-induced contribution (at weak magnetic fields).
Evidently, the interaction-induced correction to magnetoresistivity behaves as $\sim B^2T^2$ (if we apply the short-range interaction potential for inter-hole collisions, resulting in $\sigma_0(T)\sim T^2$).
It is important to note, that the sign of the second term in Eq.~\eqref{MRs3} may change depending on the particular values of the material parameters.

\subsection{The Hall effect}
Let us now discuss the interaction-induced corrections to the Hall effect.
The general expression for the Hall component of resistivity reads as
\begin{eqnarray}
\label{Hall}
\rho_{xy}(B)&=&-\rho_{yx}(B)
\\\nonumber
&=&\frac{\sigma^0_{xy}(B)+\delta\sigma_{xy}(B)}
{[\sigma^0_{xx}(B)+\delta\sigma_{xx}(B)]^2+[\sigma^0_{yx}(B)+\delta\sigma_{yx}(B)]^2}.
\end{eqnarray}
After extracting the interaction-induced correction, it takes the form:
$\rho_{xy}(B)=\rho^0_{xy}(B)+\delta\rho_{xy}(B)$, where
\begin{eqnarray}\label{Hall2}
&&\rho^0_{xy}(B)=\frac{\sigma^0_{xy}(B)}
{[\sigma^0_{xx}(B)]^2+[\sigma^0_{yx}(B)]^2},
\\\nonumber
&&\delta\rho_{yx}(B)\approx
-[\rho^0_{xy}(B)]^2\\
\nonumber
&&\times
\left\{2\frac{\sigma^0_{xy}(B)}{\sigma^0_{xx}(B)}\delta\sigma_{xx}(B)
-\left[1-\left(\frac{\sigma^0_{xy}(B)}{\sigma^0_{xx}(B)}\right)^2\right]\delta\sigma_{xy}(B)\right\}.
\end{eqnarray}
\begin{figure}[t!]
\centering
\includegraphics[width=1.02\columnwidth]{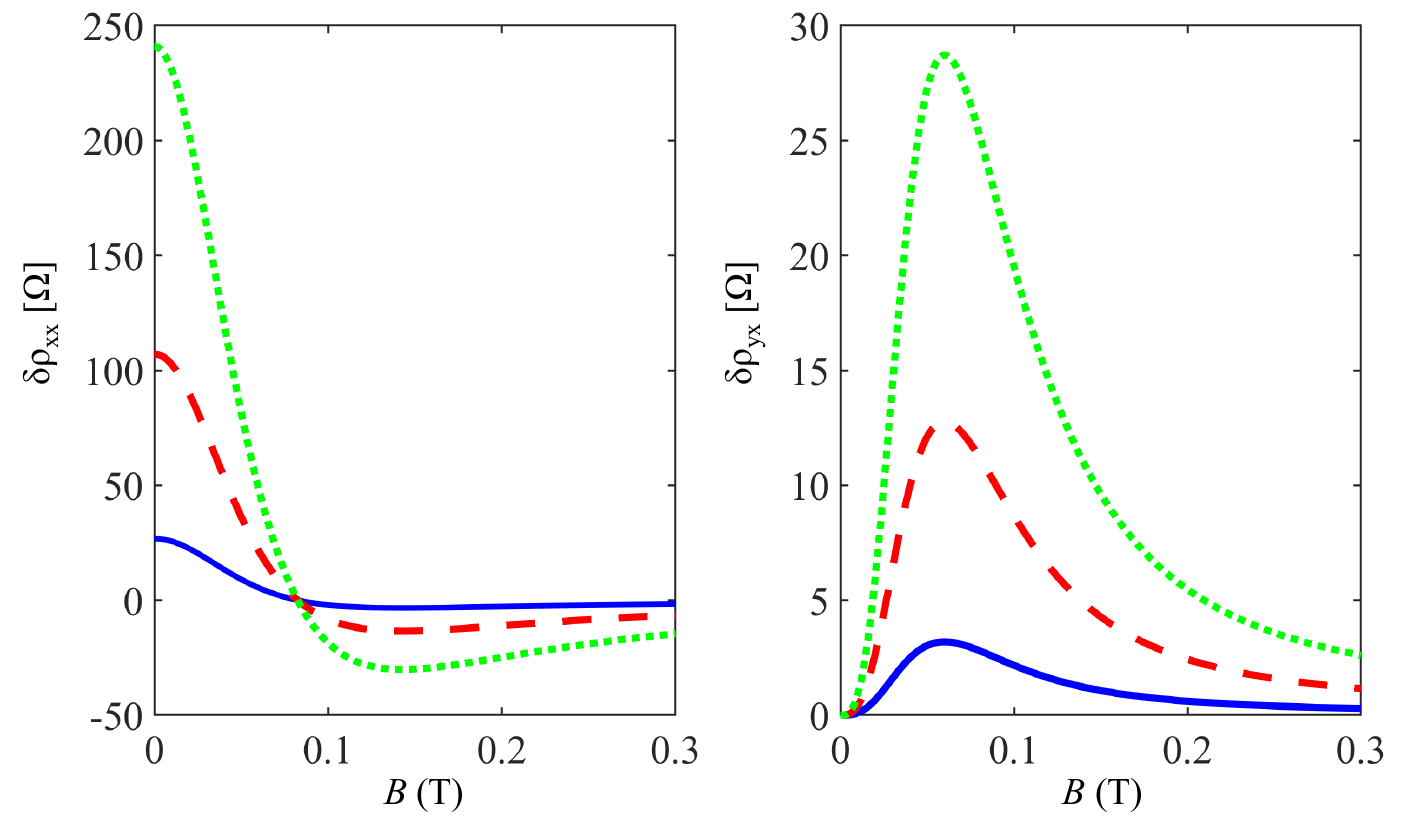}
\caption{In a mixture of massive and massless holes, the longitudinal magnetoresistance correction \(\delta\rho_{xx}\) (left) and Hall magnetoresistance correction \(\delta\rho_{yx}\) (right) vary with the magnetic field at different  temperatures: 10~K (blue), 20~K (red), and 30~K (green). The parameters are the same as in previous plots.}
\label{Fig6}
\end{figure}

As before, the interactions-induced correction to the Hall effect of interacting massless hole gas in the first order of $\delta\sigma_{0}(T)$ vanishes.
Instead, the interactions-induced correction of the massive-massless mixture at weak magnetic fields is generally nonzero:
\begin{equation}
\label{Hall3}
\delta\rho_{xy}(B)=
-\sigma_0(T)\frac{2(\beta-1)}{\beta}
\frac{\sigma^D+\beta\sigma^h}{(\sigma^D+\sigma^h)^3}
(\omega_p\tau_p-\omega_k\tau_k).
\end{equation}
It demonstrates a $\sim BT^2$ behavior.

Figure~\ref{Fig6} shows the behavior of magnetoresistivity and the Hall effect corrections with magnetic field strength in the interacting massless-massive hole mixture at differnet temperatures.
In both cases, magnetic field suppresses interaction-induced corrections to magnetoresistivity and the Hall effect accompanied by the magnetoresistivity sign change with increasing magnetic field.



\section*{Conclusions}
It is known that in the case of Galilean-invariant systems with parabolic dispersion, interaction-induced corrections to conductivity and magnetoconductivity are absent.
The violation of Galilean invariance is expected to result in finite corrections to charge transport coefficients.
Here, we developed a theory describing the interaction-mediated corrections to the magnetoconductivity and magnetoresistivity due to the interparticle interaction in p-doped 2D semiconductors with broken Galilean invariance containing two hole branches: with linear-in-momentum (Dirac-like) and parabolic dispersions.
We showed that in the regime when only the Dirac holes are present, the magnetoconductivity acquires a finite correction due to the interparticle scattering and due to the violation of Galilean invariance (while in a massive-hole subsystem any such corrections are absent).

Despite the presence of finite corrections to the massless hole magnetoconductivity,  the magnetoresistivity and the Hall effect are not affected by the interparticle correlations between massless holes, at least in the lowest order with respect to the interparticle interacting strength.
Instead, the presence of two types of holes, namely the Dirac holes and the massive holes, and an external permanent weak magnetic field provide the temperature and magnetic field-dependent contribution to the resistivity and the Hall effect.
We calculated the temperature dependence of the massless-massive hole scattering corrections for two principally different models of interparticle interacting potentials and demonstrated different temperature behavior for long-range and short-range interparticle interacting potentials. Thus, we analyzed the interparticle interacting-induced magnetotransport.


\section*{Outlook}
It should be noted that magnetoresistance may also appear due to the intervalley scattering (interparticle conversion, when Dirac particle scatters into massive hole valley and vice versa) due to the impurity potential presented in any structure.
It is known that such processes induces  finite magnetoresistance in the systems with parabolic dispersion of particles and having several subbands coupled by the particle-impurity scattering~\cite{Zaremba}. Nevertheless, the interaction-induced contributions considered in this work are strongly temperature-dependent, that opens ways to detect them in experiments.

Furthermore, we want to add that the phenomena described in this manuscript can also occur in hybrid systems like a stack of two layers with parabolic and linear carrier dispersions, as they are analogous to a HgTe system.
For example, a sheet of graphene on top of a MoS$_2$ or other TMD material constitutes an analogous system in which such interactions can take place.
Here, due to the spatial separation of the layers, there occurs a suppression of the interaction between the carriers in different layers, described by the term proportional to $\exp(-qd)$ with $d$ the distance between the layers.
This factor potentially diminishes the strength of scattering, but still can be considerable.

Finally, we want to mention that the theoretical approach developed here works for the case of degenerate hole gas far from the Dirac point.
The case of Maxwell electron and hole gases and their interaction (between Maxwell massive holes and Dirac holes) requires a separate study.

\section*{Acknowledgements}
We were supported by the National Natural Science Foundation of China (NSFC) (Grant No.~W2532001),
the Ministry of Science and Higher Education of the Russian Federation (Project FSUN-2023-0006), and the Foundation for the Advancement of Theoretical Physics and Mathematics ``BASIS''.
The work used equipment of the Center for Collective Use ``Structure, Mechanical and Physical Properties of Materials'' of the Novosibirsk State Technical University.


%
\begin{widetext}

\begin{appendix}
\section{Explicit formulas for different contributions to the electric current density}
\label{appendix}

Here, we present the explicit expressions for the electric current densities for (i) the Dirac holes in the longitudinal (x) and transverse (y) directions:
\begin{eqnarray}
\nonumber
j^D_x&=&
2\pi \textcolor{black}{\tilde g} e^2\tau_pE_x \frac{v m\mu}{(2\pi v)^2}
\int\limits_0^\infty\frac{qdq}{2\pi}
|U_{\bf q}|^2
\int\limits_{-\infty}^{+\infty}\frac{\omega^2 d\omega}{4T\sinh^2\left(\frac{\omega}{2T}\right)}
\textcolor{black}{
\frac{1}{2\pi}
\frac{\theta(4k^2q^2-(2m\omega-q^2)^2)}{\sqrt{4k^2q^2-(2m\omega-q^2)^2)}}
}
\textcolor{black}{
\frac{1}{2\pi}\frac{\theta(\varepsilon_q^2-\omega^2)}{\sqrt{(\omega^2-\varepsilon_q^2)(\varepsilon_q^2-(2\mu-\omega)^2)}}
}
\\
\label{EqMain14}
&&\times
\left[
-
\frac{4\tau_k(\mu-\omega)(\varepsilon_q^2+2\mu\omega-\omega^2)\textcolor{black}{(\tau_p\omega_p\tau_k\omega_k-1)}}{v\mu
(1+\tau_p^2\omega_p^2)
(1+\tau_k^2\omega_k^2)
}
+
\frac{4mv\tau_p(\varepsilon_q^2-\omega^2)
\textcolor{black}{(\tau_p^2\omega_p^2-1)}
}{\mu
(1+\tau_p^2\omega_p^2)^2}
\right],
\end{eqnarray}
\begin{gather}
\nonumber
j^D_y=
2\pi\textcolor{black}{\tilde g}
e^2\tau_pE_x \frac{v m\mu}{(2\pi v)^2}
\int\limits_0^\infty\frac{qdq}{2\pi}
|U_{\bf q}|^2
\int\limits_{-\infty}^{+\infty}\frac{\omega^2 d\omega}{4T\sinh^2\left(\frac{\omega}{2T}\right)}
\textcolor{black}{
\frac{1}{2\pi}
\frac{\Theta(4k^2q^2-(2m\omega-q^2)^2)}{\sqrt{4k^2q^2-(2m\omega-q^2)^2)}}
}
\textcolor{black}{
\frac{1}{2\pi}\frac{\theta(\varepsilon_q^2-\omega^2)}{\sqrt{(\omega^2-\varepsilon_q^2)(\varepsilon_q^2-(2\mu-\omega)^2)}}
}
\\
\label{EqMain15}
\times
\left[
-\frac{4\tau_k (\mu-\omega)(\varepsilon_q^2+2\mu\omega-\omega^2)\textcolor{black}{(\tau_p\omega_p+\tau_k\omega_k)}}{v\mu(1+\omega_p^2\tau_p^2)(1+\omega_k^2\tau_k^2)}
+
\frac{4m\tau_p v(\varepsilon_q^2-\omega^2)\textcolor{black}{2\tau_p\omega_p}}{\mu(1+\omega^2_p\tau_p^2)^2}
\right],
\end{gather}
and (ii) the massive holes:
\begin{eqnarray}
\nonumber
j^h_x&=&
2\pi\textcolor{black}{\tilde g}
e^2\tau_kE_x \frac{v_Fm\mu}{(2\pi v)^2}
\int\limits_0^\infty\frac{qdq}{2\pi}
|U_{\bf q}|^2
\int\limits_{-\infty}^{+\infty}\frac{\omega^2 d\omega}{4T\sinh^2\left(\frac{\omega}{2T}\right)}
\textcolor{black}{
\frac{1}{2\pi}\frac{\theta(\varepsilon_q^2-\omega^2)}{\sqrt{(\omega^2-\varepsilon_q^2)(\varepsilon_q^2-(2\mu-\omega)^2)}}
}
\textcolor{black}{
\frac{1}{2\pi}
\frac{\Theta(4k^2q^2-(2m\omega-q^2)^2)}{\sqrt{4k^2q^2-(2m\omega-q^2)^2)}}
}
\\
\label{EqMain16}
&&\times
\left[
\frac{8m\tau_k(\mu-\omega)(\frac{q^2}{2m}-\omega)
\textcolor{black}{
(\tau_k^2\omega_k^2-1)}}
{k(1+\omega_k^2\tau_k^2)^2}
-
\frac{8m^2v^2\tau_p(\frac{q^2}{2m}-\omega)(2\varepsilon_q^2\mu+\omega\varepsilon_q^2+2\mu\omega^2-\omega^3)\textcolor{black}{(\omega_p\tau_p\omega_k\tau_k-1)}}
{
k\varepsilon_q^2\mu
(1+\omega_p^2\tau_p^2)
(1+\omega_k^2\tau_k^2)}
\right],
\end{eqnarray}
\begin{eqnarray}
\nonumber
j^h_y&=&
2\pi\textcolor{black}{\tilde g}
e^2\tau_kE_x \frac{v_Fm\mu}{(2\pi v)^2}
\int\limits_0^\infty\frac{qdq}{2\pi}
|U_{\bf q}|^2
\int\limits_{-\infty}^{+\infty}\frac{\omega^2 d\omega}{4T\sinh^2\left(\frac{\omega}{2T}\right)}
\textcolor{black}{
\frac{1}{2\pi}\frac{\theta(\varepsilon_q^2-\omega^2)}{\sqrt{(\omega^2-\varepsilon_q^2)(\varepsilon_q^2-(2\mu-\omega)^2)}}
}
\textcolor{black}{
\frac{1}{2\pi}
\frac{\Theta(4k^2q^2-(2m\omega-q^2)^2)}{\sqrt{4k^2q^2-(2m\omega-q^2)^2)}}
}
\\
\label{EqMain17}
&&\times
\left[
\frac{8m\tau_k(\mu-\omega)(\frac{q^2}{2m}-\omega)\textcolor{black}{2\tau_k\omega_k}}
{k(1+\omega_k^2\tau_k^2)^2}
-
\frac{8m^2v^2\tau_p(\frac{q^2}{2m}-\omega)(2\varepsilon_q^2\mu+\omega\varepsilon_q^2+2\mu\omega^2-\omega^3)\textcolor{black}{(\tau_k\omega_k+\tau_p\omega_p)}}
{
k\varepsilon_q^2\mu(1+\omega_p^2\tau_p^2)
(1+\omega_k^2\tau_k^2)}
\right].
\end{eqnarray}
\textcolor{black}{
Here, $\tilde g=g^D_{s}g^D_{v}g^h_{s}g^h_{v}=8$, where
$g^h_{s}=2$ and $g^h_{v}=2$ are spin and valley degeneracy factors of massive holes.} These general expressions describe the currents at arbitrary values of $\omega_p\tau_p$ and $\omega_k\tau_k$, constant hole-impurity scattering times $\tau_{p(k)}$, and arbitrary (Fourier components of) hole-hole interaction potential $U_{\bf q}$.
Note that in these expressions all radicals are real valued and positive, implying positive expressions under the square roots.

\end{appendix}
\end{widetext}

\bibliography{biblio}
\bibliographystyle{apsrev4-2}


\end{document}